\documentstyle[psfig,prl,aps,multicol]{revtex}

\begin{document}
\draft                                           

\title{   One-dimensional metallic behavior of the stripe
          phase in La$_{2-x}$Sr$_x$CuO$_4$ }

\author{  Marcus Fleck, Alexander I. Lichtenstein and Eva Pavarini} 
\address{ Max-Planck-Institut f\"{u}r Festk\"{o}rperforschung, 
       Heisenbergstrasse 1, D-70569 Stuttgart, Federal Republic of Germany } 

\author{  Andrzej M. Ole\'s }
\address{ Institute of Physics, Jagellonian University, 
          Reymonta 4, PL-30059 Krak\'ow, Poland  } 

\date{\today}
\maketitle

\begin{abstract}
Using an exact diagonalization method within the dynamical mean-field 
theory we study stripe phases in the two-dimensional Hubbard model. 
We find a crossover at doping $\delta\simeq 0.05$ from diagonal stripes
to vertical site-centered stripes with populated domain walls, stable in a 
broad range of doping, $0.05<\delta<0.17$. The calculated chemical 
potential shift 
$\propto -\delta^2$ and the doping dependence of the magnetic 
incommensurability are in quantitative agreement with the experimental 
results for doped La$_{2-x}$Sr$_x$CuO$_4$. The electronic structure shows 
one-dimensional metallic behavior along the domain walls, and explains the 
suppression of spectral weight along the Brillouin zone diagonal. 
\end{abstract}


\begin{multicols}{2} 

It is commonly believed that the understanding of normal state properties 
of high-temperature superconducting cuprates (HTSC) will provide important 
clues for the understanding of superconductivity itself. The undoped 
compounds, La$_2$CuO$_4$ and YBa$_2$Cu$_3$O$_6$, are insulators and exhibit 
long-range antiferromagnetic (AF) order, which is rapidly destroyed and 
replaced by short-range AF correlations as holes are doped into the CuO$_2$ 
planes \cite{Kas98,Nie98}. Thus, one of the most important features is the 
nature of the interplay between the AF spin fluctuations and 
superconductivity. Incommensurate charge and spin order, discovered first 
in La$_{1.6-x}$Nd$_{0.4}$Sr$_x$CuO$_4$ \cite{Tra96}, suggests that the 
strong competition between hole propagation and AF order in the CuO$_2$ 
planes leads to segregation of holes in regions without AF order. These 
regions form one-dimensional (1D) substructures, so called {\it stripes\/}, 
which act as domain walls \cite{Eme93}. 
The essentialy identical momentum dependence of 
the magnetic scattering in La$_{2-x}$Sr$_x$CuO$_4$ \cite{Kas98,Yam98} 
provides evidence for the {\it stripe phases\/} in this class of materials.  

If indeed realized in a broad range of doping, the stripe phase should
have measurable consequences. Studies of 
La$_{2-x-y}$Nd$_y$Sr$_x$CuO$_4$ showed a chemical potential shift in 
underdoped and overdoped cuprates responsible for the breakdown of 
the Fermi liquid picture, a pseudogap which opens at the Fermi level 
\cite{Ino97}, and a real gap for charge excitations in the 
electronic structure around momentum $(\pi/2,\pi/2)$ \cite{Ino99,Zho99}. 
It may be expected that such puzzling features follow from 
strong Coulomb interactions at Cu ions, and it would be interesting to 
investigate whether they are fingerprints of a stripe phase and could be 
reproduced by considering a generic model for the HTSC, a two-dimensional 
(2D) Hubbard model.    
 
Stripe phases were first found in the Hartree-Fock (HF) approximation 
\cite{Zaa89}, with empty (filled by holes) domain walls in
an insulating ground state. In contrast, the calculations which 
include electron correlations indicate that the ground state of an 
AF system with strong short-range Coulomb repulsion is 
a stripe phase with populated domain walls at low doping 
\cite{Nay97,Whi98,Sei98}. Hence a partially filled band might be 
expected, and charge transport along the walls becomes possible. 
These results clearly emphasize the 
need for a reliable and controlled approximation scheme in 
order to study the physics of stripe phases. 

In this Letter we present an {\it exact} solution of the 
dynamical mean-field theory (DMFT) \cite{Geo96} equations 
for the stripe phase of the 2D Hubbard model. 
The DMFT approach allows to treat the hole correlations in a 
non-perturbative way using {\it local selfenergy\/} \cite{Met89}. 
Recently we have shown that within DMFT one obtains the correct 
dispersion and spectral weights of quasiparticle (QP) states in the 
Hubbard model at half-filling ($n=1$) \cite{Fle98}. 

Here we investigate long-range stripe order in the 2D Hubbard model at 
zero temperature. The square lattice is thereby covered by $N$ supercells 
containing $L$ sites each, and the ground state energy of the Hamiltonian, 
\begin{equation}
H=-\sum_{mi,nj,\sigma}t_{mi,nj}a_{mi\sigma }^{\dagger }a_{nj\sigma
}^{}+U\sum_{mi}n_{mi\uparrow }n_{mi\downarrow },  
\label{hubbard}
\end{equation}
has been determined using this constraint. Positions 
${\bf R}_{mi}\equiv{\bf T}_m+{\bf r}_i$ of nonequivalent sites $i=1,...,L$ 
within the unit cell $m$ are labelled by a pair of indices $\{mi\}$. 
We focus on the generic behavior of the stripe phase and thus 
restrict the hopping term to nearest-neighbors $\{mi\}$ and $\{nj\}$ only, 
$t_{mi,nj}=t$, and take a uniform on-site Coulomb interaction $U$. 
The one-particle Green's function in the stripe phase is 
given by an $(L\times L)$ matrix, $G_{ij\sigma}({\bf k},i\omega_{\nu})$, 
on the imaginary energy axis $\omega_{\nu}=(2\nu +1)\pi T$ with fictitious 
temperature $T$. It contains a site- and 
spin-dependent {\it local selfenergy} \cite{Geo96,Met89},
\begin{equation}
G_{ij\sigma}^{-1}({\bf k},i\omega_{\nu})=(i\omega_{\nu}+\mu)\delta_{ij}
-h_{ij}({\bf k})-\Sigma_{ii\sigma}(i\omega_{\nu})\delta_{ij},
\label{localg}
\end{equation}
where $\mu$ is the chemical potential, and $h_{ij}({\bf k})$ is an 
$(L\times L)$ matrix which describes the kinetic energy, 
$h_{ij}({\bf k})=\sum_n\exp(-i{\bf k}({\bf R}_{0i}-{\bf R}_{nj}))t_{0i,nj}$. 
The local Green's functions for each nonequivalent site $i$ are 
calculated from the diagonal elements of the Green's function matrix 
(\ref{localg}), $G_{ii\sigma }(i\omega_{\nu})=N^{-1}\sum_{\bf k}
G_{ii\sigma }({\bf k},i\omega_{\nu})$. 
Self-consistency of site $i$ with its effective medium requires, 
\begin{equation}
{\cal G}_{ii\sigma}^0(i\omega_{\nu})\,^{-1}
=G_{ii\sigma}^{-1}(i\omega_{\nu})+\Sigma_{ii\sigma}(i\omega_{\nu})\;, 
\label{cavity}
\end{equation}
similar to the situation in thin films \cite{Pot99}.

For the solution of the effective impurity model with hybridization 
parameters, $V_{i\sigma}(k)$, as well as diagonal energies, 
$\varepsilon_{i\sigma}(k)$, for each non-equivalent site $i$ in the stripe 
supercell we employed the exact diagonalization method of Caffarel and 
Krauth \cite{Caf94}. By fitting ${\cal G}_{ii\sigma}^0(i\omega_{\nu})$ on 
the imaginary energy axis, 
\begin{equation}
{\cal G}_{ii\sigma }^0(i\omega _\nu )\,^{-1}=i\omega _\nu +\mu
-\sum_{k=2}^{n_s}\frac{V_{i\sigma }^2(k)}
{i\omega _\nu -\varepsilon_{i\sigma }(k)}\;,  
\label{fit}
\end{equation}
the parameters of an effective DMFT-impurity cluster with 
$n_s$ sites are obtained. After solution of the effective cluster 
problem with Lanczos algorithm ($n_s \sim 8$), the local Green's 
function $G_{ii\sigma}(i\omega_{\nu})$, and local electron densities 
$n_{i\sigma}$ were determined. The self-consistency is implemented by 
extracting the new selfenergy for the next DMFT-iteration 
from Eq. (\ref{cavity}). Finally, the Green functions (\ref{localg}) 
serve to determine the spectral function,
\[
A({\bf k},\omega)=-{1\over\pi}{1\over L N}{\rm Im}\sum_{mi, nj, \sigma}
e^{-i{\bf k}({\bf R}_{mi}-{\bf R}_{nj})}G_{mi,nj,\sigma}(\omega).
\]

Below we summarize the results obtained for $U=12t$, a value 
representative for La$_{2-x}$Sr$_x$CuO$_4$ \cite{notet'}, and for a broad 
range of hole doping ($\delta=1-n$), $0.03<\delta<0.2$, where we found  
that the ground state contains populated domain walls. 
First, for $0.03<\delta\leq 0.05$, diagonal stripe supercells made out of 
pieces of site-centered vertical domain walls are stabilized by a (weak) 
charge density wave (CDW) superimposed with a spin density wave (SDW) 
along the wall. The SDW domain wall unit cells consist of four sites, 
$|0\rangle-|\uparrow\rangle-|0\rangle-|\downarrow\rangle$. 
For $\delta=0.05$ we found the somewhat lower electron densities at 
nonmagnetic ($|0\rangle$) sites ($n_0\simeq 0.848$) than the densities at 
magnetic ($|\sigma\rangle$) sites ($n_m\simeq 0.860$) with magnetic moment 
of $m\simeq 0.334$. The SDW unit cell is stabilized by only 
$\simeq 0.584$ holes, which indicates that such states are precursors of 
the undoped AF Mott insulator \cite{Kiv98}. 

\begin{figure}
\centerline{\psfig{figure=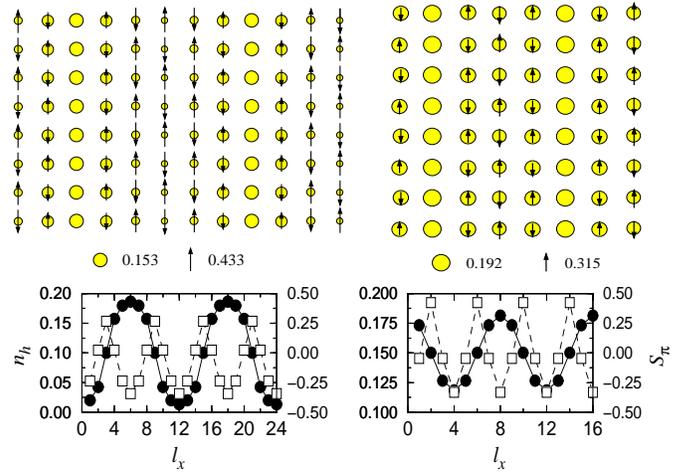,height=2.5in,width=3.6in}}
\narrowtext
\caption
{Vertical stripe phases for $\delta=1/12$ (left) and 
 $\delta=0.15$ (right) at $U=12t$. Top part shows doped hole (circles)
 and magnetization density (arrows). The spatial variation of these 
 quantities are represented by $n_h(l_x)$ (open squares) and 
 $S_{\pi}(l_x)$ (solid circles) in the lower part.}
\label{stripes}
\end{figure}

In agreement with neutron scattering experiments \cite{Tra96}, we found a 
particular stability of the vertical stripes with {\it populated domain 
walls\/}. As the most robust structure of the doped systems with 
$0.05<\delta<0.17$ we identified the site-centered stripe phase 
(Fig. \ref{stripes}). This phase is stabilized by electron correlations 
\cite{Nay97,Whi98,Sei98} and by the kinetic energy gains on the populated 
domain walls which destabilize a SDW along the walls in this doping regime. 
At higher doping $\delta>0.17$ the bond-centered stripe phase of White and 
Scalapino \cite{Whi98} is energetically favored, kinks and antikinks along 
the domain walls occur, and the stripe structure gradually melts.
 
The size of AF domains in the site-centered stripe phase shrinks with 
increasing doping for $\delta\leq 1/8$ (Fig. \ref{stripes}). For example, 
the charge unit cell contains eight (four) sites at doping $\delta=1/16$ 
($\delta=1/8$), while the electron density on the sites of the walls is 
almost constant, being $n_i\simeq 0.850$ and $n_i\simeq 0.830$, 
respectively. In agreement with the results of the numerical density matrix 
renormalization group \cite{Whi98}, the self-consistent DMFT densities in 
the stripe unit cell are characterized by much smoother variations 
than in the corresponding HF states \cite{Zaa89}. Beyond $\delta=1/8$ we 
find a lock-in effect of the same structure with a charge (magnetic) unit 
cell consisting of four (eight) sites, and the doped hole density, 
$n_h(l_x)=1-\langle n_{(l_x,0),\uparrow}+n_{(l_x,0),\downarrow}\rangle$, 
increasing faster within the AF domains than on the wall sites 
(Fig. \ref{stripes}). 
The magnetic domain structure is best described by the modulated 
magnetization density,
%
$S_{\pi}(l_x)=L_y^{-1}\sum_{l_y}(-1)^{l_x+l_y}
\frac{1}{2}\langle n_{(l_x,l_y),\uparrow}-n_{(l_x,l_y),\downarrow}\rangle$,
%
projected on the direction perpendicular to the wall \cite{Whi98}. 

\begin{figure}
\centerline{\psfig{figure=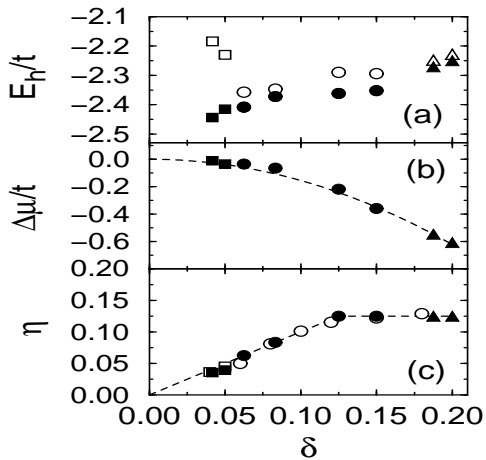,height=2.5in,width=2.5in}}
\narrowtext
\caption
{Evolution of the stripe phases with doping $\delta$ ($U=12t$): 
 (a) Energy per doped hole $E_h/t$ in diagonal SDW stripes 
     (squares), and vertical (circles) site-centered and  
     bond-centered (triangles) stripes; filled (empty) symbols show
     the energy in the ground (excited) states. 
 (b) Shift of the chemical potential $\Delta\mu /t$ (filled symbols) 
     and the quadratic fit $\Delta\mu /t=a\delta^2$, with $a=-15.57$ 
     (dashed line).
 (c) Shift $\eta$ of the maxima of 
     $S({\bf Q})$ in: diagonal SDW stripes with 
     ${\bf Q}=[(1\pm \sqrt{2}\eta)\pi,(1\pm \sqrt{2}\eta)\pi]$ 
     (filled squares), and vertical site-centered (filled circles) and 
     bond-centered (filled triangles) stripes, 
     both with ${\bf Q}=[(1\pm 2\eta)\pi,\pi]$. 
     Empty symbols show the data points of Yamada {\it et al.\/} 
     \protect\cite{Yam98} (circles) and 
     Wakimoto {\it et al.\/} \protect\cite{Wak99} (squares). }
\label{doping}
\end{figure}

The stability of the above stripe phases is investigated by the ground 
state energy normalized per density of doped holes, 
$E_h=[E_0(\delta)-E(0)]/\delta$, 
where $E_0(\delta)$ is the ground state energy at doping $\delta$. 
The energy $E_h/t$ is a monotonically increasing function of doping (Fig. 
\ref{doping}(a)), showing that the different stripe phases are stable 
against macroscopic phase separation. From our results we conclude that 
short-range Coulomb repulsion suffices to obtain populated domain walls 
over a wide range of doping. The diagonal stripes, stable at 
low doping $\delta<0.06$, are followed by vertical site-centered stripes 
which compete with bond-centered stripes, both having a 
considerably larger energy gain than homogenous phases, 
such as spin spirals. We have verified that the kinetic energy is 
gained mainly on the sites which belong 
to the domain walls. Such energy gains are larger at the nonmagnetic 
domain walls (Fig. \ref{doping}(a)) than on the magnetic sites of 
bond-centered domain walls. At the same time the small energy difference 
between these both quite different states indicates a strong tendency 
towards transverse stripe fluctuations which might enhance superconducting 
correlations in the ground state \cite{Kiv98}.

We have found that the chemical potential shifts downwards with
hole doping, $\Delta\mu\propto -\delta^2$ (Fig. \ref{doping}(b)), in
agreement with the experimental results of Ino {\it et al.\/} \cite{Ino97},  
and with the Monte-Carlo simulation of a 2D Hubbard model \cite{Fur93}. 
Therefore, the charge susceptibility is enhanced towards $\delta \to 0$, 
reproducing a universal property of the Mott-Hubbard metal-insulator 
transition \cite{Fur93}.

As both the bond-centered and site-centered stripe phases have the same 
size of the magnetic unit cell, they give the same pattern in neutron 
scattering and are thus indistinguishable experimentally. 
The neutron scattering structure factor $S({\bf Q})$ in the stripe phase 
has the maxima shifted away from the $M=(\pi,\pi)$ point 
to ${\bf Q}=[(1\pm 2\eta_{\rm vert})\pi,\pi]$ points for the structures of 
Fig. \ref{stripes} [and to ${\bf Q}=[\pi,(1\pm 2\eta_{\rm vert})\pi]$ for 
equivalent horizontal stripes]. The present calculations give a linear 
dependence $\eta_{\rm vert}=\delta$ for $\delta\leq 1/8$ and 
$\eta_{\rm vert}=1/8$ for $\delta>1/8$ (Fig. \ref{doping}(c)). 
Such a behavior was observed by Yamada {\it et al.} \cite{Yam98}, and 
indicates a unique stability of populated domain walls in the stripe phase. 
The correlations included within the DMFT and its capability 
to describe the Mott-Hubbard metal-insulator transition play thereby a 
crucial role, as other filling and periodicity of the stripe phase are 
found in HF calculations \cite{Zaa89}. 
Also the points found at low doping (Fig. \ref{doping}(c)) 
corresponding to the diagonal stripe structures 
${\bf Q}=[(1\pm 2\eta_{\rm diag})\pi,(1\pm 2\eta_{\rm diag})\pi]$ agree 
perfectly well with the recent neutron experiments of 
Wakimoto {\it et al.} \cite{Wak99}. 
We find $\eta_{\rm diag} \simeq \delta/\sqrt{2}$, where the factor 
$1/\sqrt{2}$ is due to the rhombic lattice constant in diagonal 
stripe structures as suggested by experiment \cite{Wak99}.  
This results in the relation 
$\eta_{\rm diag}=\eta_{\rm vert}/\sqrt{2}$, and the linear dependence 
$\eta_{\rm vert}\simeq\delta$ holds also for $\delta < 0.06$ 
(Fig. \ref{doping}(c)). 
Furthermore, our calculations predict vertical SDW domain wall unit 
cells in the diagonal stripe phase. 
Thus additional elastic magnetic superlattice peaks should be visible 
in neutron scattering experiments 
around ${\bf Q}=[(1+2\eta_{\rm diag})\pi/2,(1+2\eta_{\rm diag})\pi]$ 
with a weight smaller by a factor $\sim 3.7$, if 
such phases do exist in heavily underdoped 
La$_{2-x-y}$Nd$_{y}$Sr$_x$CuO$_4$. 

\begin{figure}
\centerline{\psfig{figure=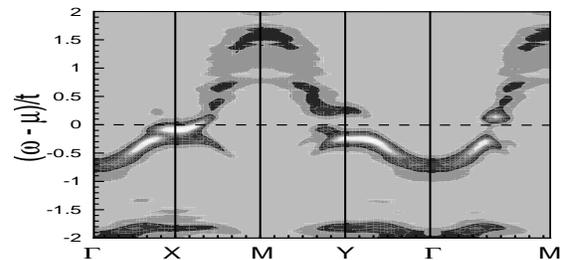,height=1.55in,width=3.2in}}
\narrowtext
\caption
{Spectral function $A({\bf k},\omega)$ along the main directions of the 
 2D BZ of the stripe phase at $\delta=1/12$ with $U=12t$.}
\label{under}
\end{figure}
\begin{figure}
\centerline{\psfig{figure=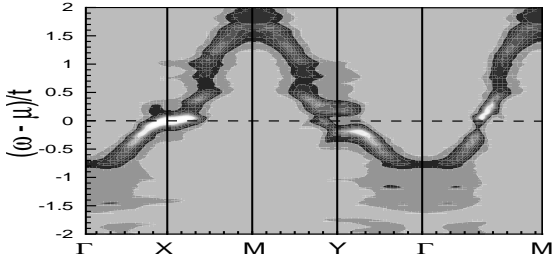,height=1.55in,width=3.2in}}
\narrowtext
\caption
{$A({\bf k},\omega)$ as in Fig. \protect\ref{under}
 but at $\delta=0.15$. }
\label{over}
\end{figure}

Let us focus on the  spectral function $A({\bf k},\omega)$ of the 
stripe phases shown in Fig. \ref{stripes}. 
The photoemission $\omega \leq \mu$ spectra consist of the 
lower Hubbard band at an energy $\omega - \mu \sim -4.8t$ 
and low-energy states, well separated 
from the Hubbard band, extending over an energy range of $\sim 2t$. 
The QP states known from the dispersion of a single 
hole in the $t$-$J$ model survive also in the stripe 
phase up to $\delta=0.15$ and are characterized by a considerable 
spectral weight and a bandwidth $\sim 2J$ (here $J/t=4t/U=1/3$) 
\cite{Fle98} (Figs. \ref{under} and \ref{over}). 
Due to the stripe structure we find that the directions 
$\Gamma-X$ [$X=(\pi,0)$] and $\Gamma-Y$ [$Y=(0,\pi)$] are 
nonequivalent. 

At $\delta=1/12$ we observe a {\it pseudogap} at the $X$ point 
(Fig. \ref{under}). The QP weight there is composed out of QP states 
originating from the dressing of a moving hole by quantum fluctuations 
in an AF background and localized states from the 1D electronic 
structure of the site-centered stripe phase \cite{notex}. 
This superposition of QP weight explains the {\it flat band} around 
the $X$ point and the Fermi level crossing at $(\pi,\pi/4)$ 
observed in recent angle-resolved photoemission (ARPES) 
experiments \cite{Ino99,Zho99}. 
On the contrary, the QP's originating from the 
1D features of the stripe phase, are not seen in ARPES around 
the $Y$ point, as the structure factor vanishes \cite{Mar99}, and 
one only resolves the spin-polaron QP with 
dispersion $\sim 2J$. 

As the most spectacular result, a {\it gap\/} for charge excitations 
opens in the underdoped regime at the Fermi energy around 
$S=(\pi/2,\pi/2)$ (Fig. \ref{under}). 
Little or no spectral weight is found at momenta 
${\bf k}=(\pi/4,\pi/4)$ and ${\bf k}=(0,\pi/4)$ where, notably, 
the 1D electronic 
structure should show Fermi level crossings \cite{notex}. 
This behavior agrees quantitatively with the ARPES measurements 
on La$_{2-x}$Sr$_x$CuO$_4$ \cite{Ino99} and 
La$_{1.28}$Nd$_{0.6}$Sr$_{0.12}$CuO$_4$ \cite{Zho99}.

In order to understand the gap structure we calculated 
the electronic structure of 
$H=-t\sum_{<ij>,\sigma} a_{i\sigma}^{\dagger }a_{j\sigma}^{} + 
\sum_i U_i S^z_i$, where $U_i$ are site-dependent spin potentials 
in the stripe supercell. Local energy contributions 
$V_i \equiv U_i |\langle S^z_i \rangle|$ are treated as parameters, 
where $V_{0}=0$ on the domain wall.
We find a condition for vanishing 
photoemission weight at momentum ${\bf k}=(\pi/4,\pi/4)$, 
$V_{0+2l_x}\geq 2V_{0+l_x}$, which is satisfied by the present 
self-consistent DMFT spin densities, with 
$V_{0+2l_x}\simeq 2.07t$ and 
$V_{0+l_x}\simeq 0.99t$ (Fig. \ref{stripes}). 
The strong renormalization of $V_i$ in DMFT is due to 
charge fluctuations and demonstrates that local correlations play a 
crucial role in understanding the ARPES spectra of 
HTSC \cite{Ino99,Zho99}.

The ARPES spectral weight at $\delta=1/12$ around 
$\omega - \mu \sim -2t$ entirely originates from bands of the stripe 
supercell (Fig. \ref{under}).
As expected, spectral weight is transferred with increasing doping from 
that energy region to the inverse photoemission 
$\omega>\mu$ (Fig. \ref{over}). Finally, we observed that the gaps 
at the $Y$ and $S$ point are gradually 
filled by spectral weight as the stripe order melts. 

Summarizing, we have shown that vertical stripe phases with populated 
domain walls are robust structures in a broad range of $\delta$. 
Their spectral properties show an interesting superposition of the 
QP's known from doped 2D antiferromagnets with a 1D metallic behavior. 
Such experimental features as:
 (i)   the chemical potential shift $\Delta\mu\propto -\delta^2$, 
 (ii)  the incommensurability of spin fluctuations, and 
 (iii) the gradual disappearence of the photoemission flat band and the
       pseudogap (gap) at the $X$ ($S$) point, 
find a natural explanation and accompany a gradual crossover from the 
stripe phases into a (strongly correlated) Fermi liquid with increasing 
doping.  

We thank O. K. Andersen, B. Keimer, T. M. Rice, and 
J. Zaanen for stimulating discussions. 
A.M.O. acknowledges the support by the Committee 
of Scientific Research (KBN) of Poland, Project No. 2~P03B~175~14. 


\end{multicols} 

\end{document}